# Conception of Brownian coil

Author: Zhang Jiayuan

Affiliation: Affiliated High School of Northwestern Polytechnical University**Abstract:**
This article proposes a conception of Brownian coil. Brownian coil is a tiny coil with the same size of pollen. Once immersed into designed magnetic field and liquid, the coil will be moved and deformed macroscopically, due to the microscopic thermodynamic molecular collisions. Such deformation and movement will change the magnetic flux through the coil, by which an ElectroMotive Force (EMF) is produced. In this work, Brownian heat exchanger and Brownian generator are further designed to transform the internal energy of liquid into other form: 1) the internal energy of the resistance; 2) the constant electric energy. The two forms accord with the Clausius' statement and Kelvin's statement respectively. If the ideas can be realized, the second law of thermodynamics and the second kind of perpetual-motion machine should be understood again.

**Keywords:** Second law of thermodynamics; Second kind of perpetual-motion machine; Brownian coil;## 1 Introduction

The Second law of thermodynamics [1] is a basic physical principle. Its Clausius' statement is: it is impossible to transfer heat from cold object to hot object without causing other changes; and the Kelvin's statement is: it is impossible to make such a heat engine in cycle movement, which transforms heat from a single heat source to work completely, without causing other changes. The second law of thermodynamics gives death sentence to the second kind of perpetual-motion machine. However, the second law of thermodynamics still lacks of strict proof, as a result, some objections emerged, such as Maxwell demon [1], as is shown in Figure 1. What should be pointed out is: although the Maxwell demon has been proved unachievable, its proposal leads to a new opinion: the transmission of information requires energy. Therefore, it is not important whether perpetual-motion machine can be made. The new thought, inspiration and fun it triggers is most valuable.

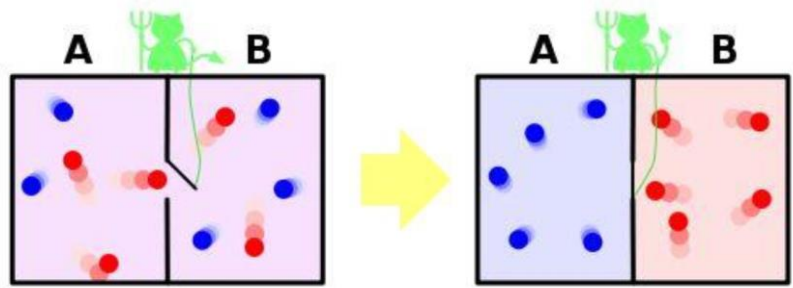

**Figure 1 Maxwell demon**

Inspired by Brownian movement, a new conception is proposed. Brownian movement [2] is the irregular motion caused by the imbalance impact of liquid particles, which is first observed by botanist R. Brown in soliquoid of pollen in 1827, as shown in Figure 2. In Brownian movement, the microscopic thermal motion of water molecules is transformed into the macroscopic movement of the pollen, in other word, the internal energy of water is converted to the kinetic energy of pollen. This article is not an attempt to collect the internal energy directly, alternatively, collect the energy of the pollen.

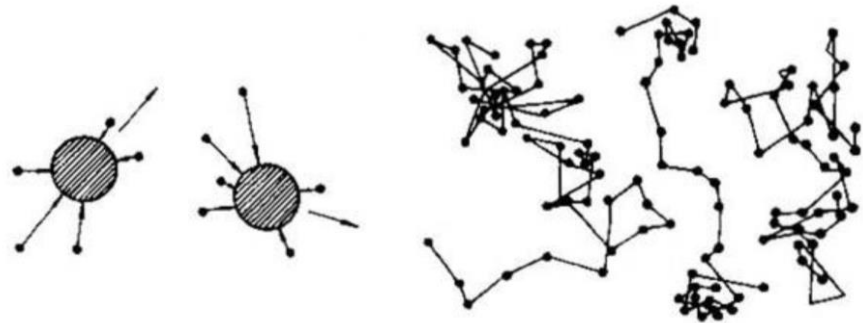

(a) Micro particles are impacted by liquid molecules    (b) Brownian movement
**Figure 2 Brownian movement**

## 2 Brownian Unit

Brownian coil is a soft and tiny metal wire with same size of pollen; immersed into appropriate magnetic field and liquid, the coil becomes a Brownian Unit; Basement membrane in Figure 3 is a membrane structure (Similar to the photosynthetic lamellar membrane carrying enzymes in cyanobacteria) on which large number of Brownian units are installed.

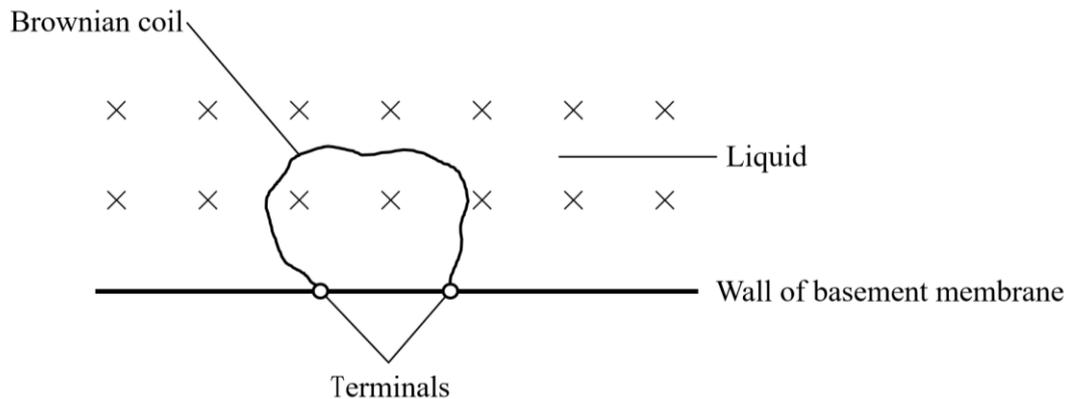

**Figure 3 Brownian coil and Brownian Unit**

Due to the thermal motion of the liquid molecules and the small size of the Brownian coil, the imbalance of the coil induced by the impact of liquid molecules is evident. Since the coil is extremely soft, the shape of the coil is easy to be changed, hence the effective area and magnetic flux will be changed and pulsing EMF is produced (assuming that the time of the deformation process is very short, while the time interval between two deformations is much longer). Such EMF is referred to *Brownian EMF*, as is shown in Figure 4 (As the Brownian coil is in reciprocating movement, the Brownian EMF will be inversed randomly and inevitably. The clockwise current can be defined as positive). Brownian EMF is macroscopic, weak, pulsed and irregular.

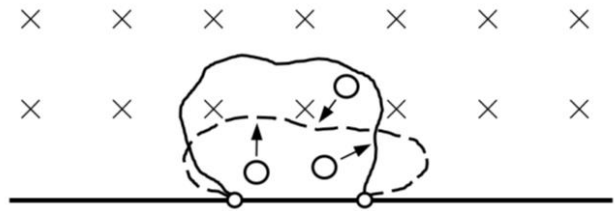

(a) Positive Brownian EMF

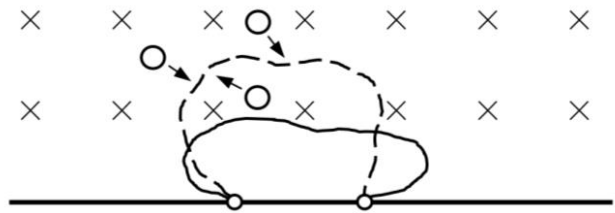

(b) Negative Brownian EMF

**Figure 4 Brownian EMF**

In light of this, the flow chart of the function of Brownian unit is demonstrated in Figure 5. What should be emphasized is: although Brownian coil is tiny, it's still macroscopic object in contrast of real microscopic object such as molecules.

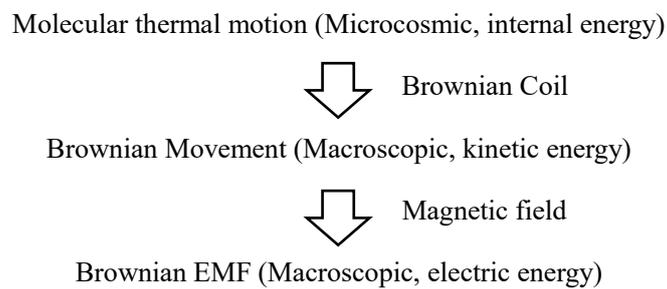

**Figure 5 Flow chart of Brownian unit**

## 3 Brownian heat exchanger

Further, a resistances is connected on the terminals of each Brownian unit to form a Brownian heat exchanger unit, as shown in Figure 6. The Brownian heat exchanger unit can directly convert the internal energy of the liquid to the internal energy of the resistance in the form of Joule heat, regardless of the temperature difference between the liquid and the resistance. However, although the produced Joule heat is macroscopic, it's still too tiny to measure and in itself a pulsing output. To solve the problem, large number of Brownian heat exchanger units should be installed on the basement membrane to construct a Brownian heat exchanger. Hence, constant and powerful heat output can be realized.

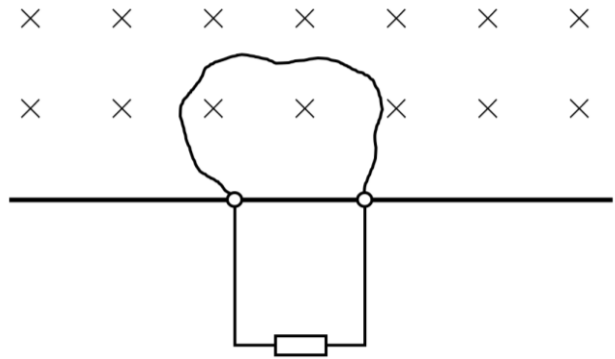

**Figure 6 Brownian heat exchanger unit**

Based on above principle, the flow chart of Brownian heat exchanger is demonstrated in Figure 7.

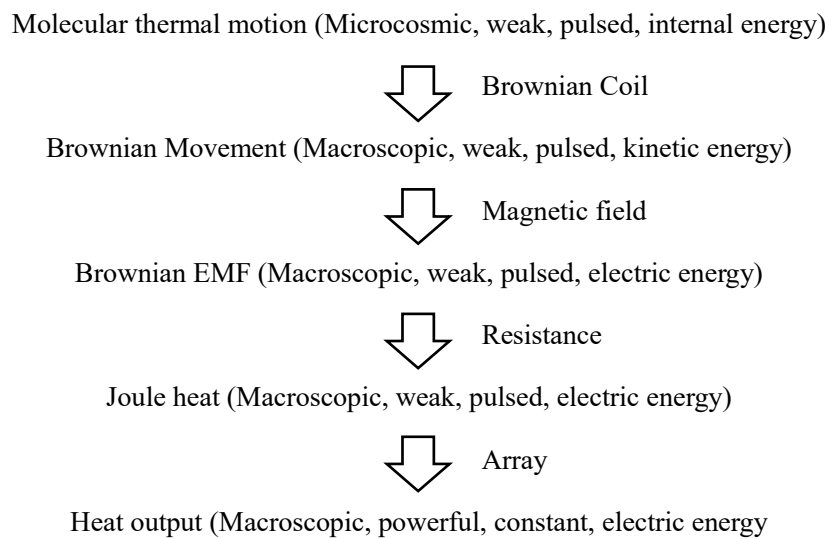

**Figure 7 Logic flow chart of Brownian heat exchanger**

Once the idea is realized, it is possible to transfer heat from colder object to hotter object without causing other changes.

## 4 Brownian generator

Although macroscopic EMF can be produced by Brownian Unit, such EMF is still weak and pulsed. To produce constant and powerful electricity, irregular Brownian EMF must be reordered. In light of this, Brownian generator is designed: a diode is connected on one side of the terminals for each Brownian unit (all the Brownian units must keep same direction), then, connect all Brownian units to mainline, as shown in Figure 8.

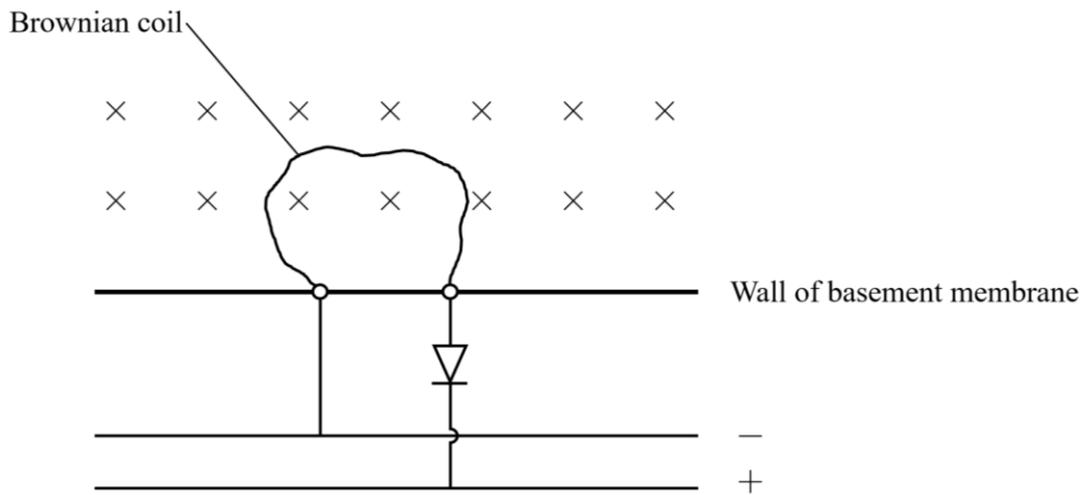

**Figure 8 Installation of diode in Brownian generator**

What can be concluded by analyzing the movement of Brownian coil is: when the effective area of the coil increases, positive EMF is produced, hence the diode is connected and Brownian unit is incorporated into the main circuit, as shown in figure 9; on contrary, when the effective area of the coil is decreased, negative EMF is produced, thus the diode is not connected and Brownian unit is offline, as shown in figure 10. In this way, although Brownian EMF for a single unit is weak, due to the large total number, considerable power output is not impossible.

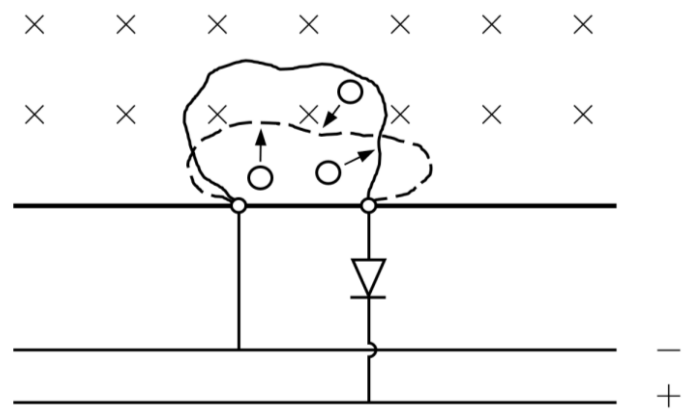

**Figure 9 An increase in effective area**

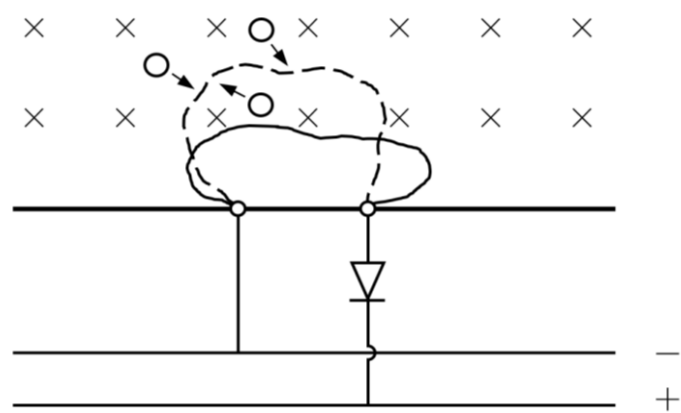

**Figure 10 An decrease in effective area**

According to the above principles, the function of Brownian generator can be summarized into a flow chart shown in Figure 11:

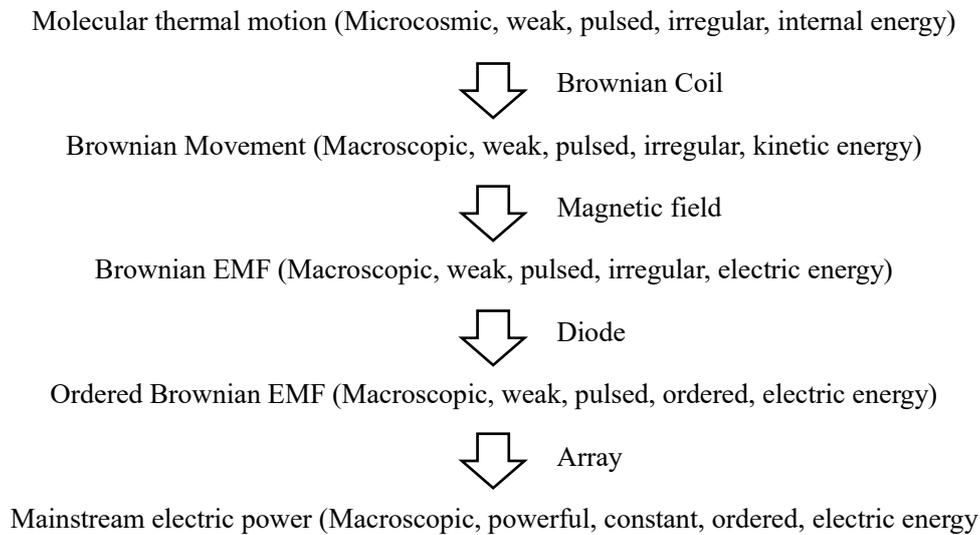

Figure 11 Logic flow chart of Brownian generator

Once the idea of Brownian generator can be realized, it is possible to make such a heat engine in cycle movement, which transforms heat from a single heat source to work completely, without causing other changes.

## 5 Conclusion

This work proposes a conception of Brownian coil. Brownian coil is a tiny coil with same size of pollen. Once immersed into designed magnetic field and liquid, due to molecular collisions, the coil will experience macroscopic movement and deformation. Such deformation and movement will change the magnetic flux through the coil, which produces EMF. Resistances are installed for large number of Brownian units, then a Brownian heat exchanger is obtained. By adopting such heat exchanger, it is hopefully that the internal energy of the liquid can be converted to the internal energy of the resistance (regardless of the temperature difference), which means it's possible to transfer heat from colder object to hotter object without triggering other changes. Further, diodes are installed on large number of Brownian units, then Brownian generator is developed. A constant and powerful output is possible by using such devise. The generator implies that it is possible to make such a heat engine in cycle movement, which transforms heat from a single heat source to work completely, without causing other changes. If both conceptions can be realized, the second law of thermodynamics and the second kind of perpetual-motion machine should be understood again.

## Acknowledgements

Thanks for Su Weidong, associate professor of Peking University, bachelor Zhang Shengqi, and Yuan Jingyang in Affiliated High School of NWPU. Thanks for their extensive and meaningful discussion.